\documentclass[10pt]{article}
\usepackage{amssymb, amsmath}
\usepackage{float,epsfig}
\usepackage{color}
\usepackage[all]{xy}

\begin{document}

\newtheorem{theorem}{\bf Theorem}[section]
\newtheorem{proposition}[theorem]{\bf Proposition}
\newtheorem{definition}[theorem]{\bf Definition}
\newtheorem{corollary}[theorem]{\bf Corollary}
\newtheorem{example}[theorem]{\bf Example}
\newtheorem{exam}[theorem]{\bf Example}
\newtheorem{remark}[theorem]{\bf Remark}
\newtheorem{lemma}[theorem]{\bf Lemma}
\newcommand{\nrm}[1]{|\!|\!| {#1} |\!|\!|}

\newcommand{\ba}{\begin{array}}
\newcommand{\ea}{\end{array}}
\newcommand{\von}{\vskip 1ex}
\newcommand{\vone}{\vskip 2ex}
\newcommand{\vtwo}{\vskip 4ex}
\newcommand{\dm}[1]{ {\displaystyle{#1} } }

\newcommand{\be}{\begin{equation}}
\newcommand{\ee}{\end{equation}}
\newcommand{\beano}{\begin{eqnarray*}}
\newcommand{\eeano}{\end{eqnarray*}}
\newcommand{\inp}[2]{\langle {#1} ,\,{#2} \rangle}
\def\bmatrix#1{\left[ \begin{matrix} #1 \end{matrix} \right]}
\def \noin{\noindent}
\newcommand{\evenindex}{\Pi_e}



\def \R{{\mathbb R}}
\def \C{{\mathbb C}}
\def \K{{\mathbb K}}
\def \J{{\mathbb J}}
\def \Lb{\mathrm{L}}

\def \T{{\mathbb T}}
\def \Pb{\mathrm{P}}
\def \N{{\mathbb N}}
\def \Ib{\mathrm{I}}
\def \Ls{{\Lambda}_{m-1}}
\def \Gb{\mathrm{G}}
\def \Hb{\mathrm{H}}
\def \Lam{{\Lambda_{m}}}
\def \Qb{\mathrm{Q}}
\def \Rb{\mathrm{R}}
\def \Mb{\mathrm{M}}
\def \norm{\nrm{\cdot}\equiv \nrm{\cdot}}

\def \P{{\mathbb P}_m(\C^{n\times n})}
\def \A{{{\mathbb P}_1(\C^{n\times n})}}
\def \H{{\mathbb H}}
\def \L{{\mathbb L}}
\def \G{{\mathcal G}}
\def \S{{\mathbb S}}
\def \sigmin{\sigma_{\min}}
\def \elam{\sigma_{\epsilon}}
\def \slam{\sigma^{\S}_{\epsilon}}
\def \Ib{\mathrm{I}}
\def \Tb{\mathrm{T}}
\def \d{{\delta}}

\def \Lb{\mathrm{L}}
\def \N{{\mathbb N}}
\def \Ls{{\Lambda}_{m-1}}
\def \Gb{\mathrm{G}}
\def \Hb{\mathrm{H}}
\def \Delta{\triangle}
\def \Rar{\Rightarrow}
\def \p{{\mathsf{p}(\lam; v)}}

\def \D{{\mathbb D}}

\def \tr{\mathrm{Tr}}
\def \cond{\mathrm{cond}}
\def \lam{\lambda}
\def \sig{\sigma}
\def \sign{\mathrm{sign}}

\def \ep{\epsilon}
\def \diag{\mathrm{diag}}
\def \rev{\mathrm{rev}}
\def \vec{\mathrm{vec}}

\def \sk{\mathsf{skew}}
\def \sy{\mathsf{sym}}
\def \en{\mathrm{even}}
\def \odd{\mathrm{odd}}
\def \rank{\mathrm{rank}}
\def \pf{{\bf Proof: }}
\def \dist{\mathrm{dist}}
\def \rar{\rightarrow}

\def \rank{\mathrm{rank}}
\def \pf{{\bf Proof: }}
\def \dist{\mathrm{dist}}
\def \Re{\mathsf{Re}}
\def \Im{\mathsf{Im}}
\def \re{\mathsf{re}}
\def \im{\mathsf{im}}

\def \sym{\mathsf{sym}}
\def \sksym{\mathsf{skew\mbox{-}sym}}
\def \odd{\mathrm{odd}}
\def \even{\mathrm{even}}
\def \herm{\mathsf{Herm}}
\def \skherm{\mathsf{skew\mbox{-}Herm}}
\def \str{\mathrm{ Struct}}
\def \eproof{$\blacksquare$}
\def \proof{\noin\pf}

\def \bS{{\bf S}}
\def \cA{{\cal A}}
\def \E{{\mathcal E}}
\def \X{{\mathcal X}}
\def \F{{\mathcal F}}
\def \tr{\mathrm{Tr}}
\def \range{\mathrm{Range}}

\def \pal{\mathrm{palindromic}}
\def \palpen{\mathrm{palindromic~~ pencil}}
\def \palpoly{\mathrm{palindromic~~ polynomial}}
\def \hodd{H\mbox{-}\odd}
\def \heven{H\mbox{-}\even}


\title{Laplacian matrices of weighted digraphs represented as
quantum states\thanks{This work is supported by CSIR (Council of Scientific and Industrial Research) Grant No. 25(0210)/13/EMR-II, New Delhi, India. }}

\author{ Bibhas Adhikari\thanks{Department of Mathematics,
IIT Kharagpur, India, E-mail:
bibhas@maths.iitkgp.ernet.in} \,\, Subhashish Banerjee\thanks{Department of Physics,
IIT Jodhpur, India, E-mail: subhashish@iitj.ac.in} \,\, Satyabrata Adhikari\thanks{Department of Mathematics, BIT Mesra, India, E-mail: 	tapisatya@gmail.com}  \and and Atul Kumar\thanks{Department of Chemistry,
IIT Jodhpur, India, E-mail: atulk@iitj.ac.in}  }
\date{}

\maketitle \thispagestyle{empty}

\begin{abstract} Graph representation of quantum states is becoming an increasingly important area of research to investigate combinatorial properties of quantum states
which are nontrivial to comprehend in standard linear algebraic density matrix based approach of quantum states. In this paper, we propose a general weighted directed
graph framework for investigating properties of a large class of quantum states which are defined by three types of Laplacian matrices associated with such graphs. We
generalize the standard framework of defining density matrices from simple connected graphs to density matrices using both combinatorial and signless Laplacian matrices
associated with weighted directed graphs with complex edge weights and with/without loops. We also introduce a new notion of Laplacian matrix which we call signed
Laplacian matrix associated with such graphs. We produce necessary and/or sufficient conditions for such graphs to represent pure
and mixed quantum states. Using these criteria we finally determine the graphs whose corresponding density matrices represent entangled pure states which are well-known
and important for quantum computation applications. It is important to observe that all these entangled pure states share a common combinatorial structure.
\end{abstract}


\section{Introduction}

Quantum mechanics deals with states living in the Hilbert space, allowing for linear superpositions to be built up, a facility of immmense importance for harnessing the
power of quantum mechanics but at the same time making it computationally a formidable task. This can be most easily appreciated by
considering entanglement \cite{Einstein,wooter98} in higher dimensions as well as in multi-partite systems \cite{multient},
all mathematically and computationally very formidable tasks. Any tool that would aid in this regard would be very welcome.

In combinatorics a graph is a collection of vertices and edges which link two vertices. A digraph is a graph consisting directed edges. A loop is an edge that joins a vertex with itself. A graph is said to be a weighted graph if each edge is assigned a nonzero number which is called the weight of the corresponding edge. The theory of graphs is a well-developed
mathematical theory that has found many applications in diverse areas, such as the spectrum of a discrete Schr$\ddot{o}$dinger operator in a uniform, periodic magnetic
field \cite{sunada}. Graphs have, by their very construction, the inherent feature of visualization. A pertinent question to ask is whether graphical representation of
quantum states can be made? This would enable the incorporation of the mathematical machinery of graphs into the problems of quantum mechanics and at the same time bring
in the attractive feature of visualization of quantum states.

Attempts have been made to realize density matrix representation of a quantum state by defining a matrix associated with a graph. In this case, the graph is called the
graph representation of a quantum state. This idea was first introduced in \cite{BrGhSe06} by considering a combinatorial Laplacian matrix associated with an unweighted
undirected graph (simple graph). It was further extended in \cite{HasJoa} for weighted graphs. 
A criteria of separability of multipartite states represented by the
combinatorial Laplacian matrices of simple graphs have also developed in \cite{WChiWa}. Recently, local unitary
transformations on a density matrix obtained by signless Laplacian matrix associated with a simple graph has been established as a combinatorial operation which is known
as switching of a graph in \cite{DuAdBa}. A combinatorial
operation has also been introduced for density matrices defined by Laplacian matrices associated with simple graphs in \cite{DuAdBaSr} that act as an entanglement
generator for mixed states arising from partially symmetric graphs.

In this paper, we use both combinatorial and signless Laplacian matrices to define density matrices associated with a weighted digraph having complex edge weights and with
or without loops. We also introduce a matrix, which we call signed Laplacian matrix associated with a weighted digraph having loops with both positive and negative
weights. In order to relate the topological structure of a weighted digraph and properties of the density matrices defined by these
Laplacian matrices, we investigate the zero eigenvalues of these matrices associated with such graphs. This leads to a classification of
graphs which always provide density matrices representing pure quantum states, and those which determine mixed quantum states. Indeed, since the number of mixed states
are significantly larger than the number of pure states, emphasis is given on identifying graphs which can produce important classes of pure states. We also provide
graphs which identify a number of well-known entangled pure states by using the Laplacian matrices associated with such graphs.
A state vector of an entangled state cannot be expressed as a tensor product of other state vectors. These states play an important role in different tasks of quantum
information theory. We observe that in the graph
representation of entangled pure states, all the weighted edges are clustered in a subgraph which forms a completely connected graph and the weight of the loops
attached at each of the vertices is $-(m-2)$ where $m$ is the number of vertices involved in the complete subgraph.  This should be of interest to the quantum
information community.

  The plan of the paper is as follows.  In section 2, we define the required terminologies of graph theory and in particular, we investigate the Laplacian matrices and
  Laplacian spectra  associated with a weighted digraph with or without   loops. In section 3, we define the density matrix associated with a weighted digraph and
  classify the graphs which represent pure and mixed states. Finally, we provide graphs which  define density matrices of entangled pure states.

\section{Weighted digraphs with/without loops and its Laplacian spectra}

Let $G=(V, E)$ be a graph with the vertex set $V = \{1, 2, \hdots, n\}$ and edge set $E \subseteq V\times V.$ A directed graph or digraph $G$ is a graph with a function assigning to each edge an ordered pair of vertices. The first vertex of the ordered pair is called the initial vertex of the edge, and the second is the terminal vertex; together, they are the endpoints. Thus each edge of a digraph is directed; and an undirected edge can be considered as both way directed. A weighted graph $G$ is a graph with a function $w: E\rightarrow \C,$ defined by $w(i,j) = w_{ij}, 1\leq i, j \leq n,$ where $\C$ is the set of all complex numbers. If $|w_{ij}|=1$ for all $(i,j)\in E$, such a graph is called gain graph \cite{Nathan}. The function $w$ is called an edge-weight function and $w_{ij}$ is called the weight of $(i,j)\in E.$ An unweighted graph can be considered as an edge-weighted graph with weight function $w_{ij} =1$ if $w_{ij}\neq 0$. A weighted digraph is a graph which is both weighted and directed. However, as mentioned, an unweighted undirected graph also can be considered as weighted digraph with constant weight function with weight of any edge is $1$ and the edges are both way directed. Thus, from now onward a graph always is a weighted digraph unless otherwise mentioned.

\subsection{Weighted digraphs with or without loops having nonnegative weights}

First, we consider weighted digraphs having loops (at least one loop and maximum
one loop at a vertex) with nonnegative real weights. The adjacency matrix $A(G)=(a_{ij})$ associated with $G$ is defined as
$$a_{ij} = \left\{
  \begin{array}{ll}
  w_{ij}, & \hbox{if $(i, j)\in E$;} \\
 \overline{w}_{ij}, & \hbox{if $(j, i)\in E$;} \\
r_i , & \hbox{if $(i, i)\in E$;} \\
0, & \hbox{otherwise}
  \end{array}
 \right.
$$ where $0\leq r_i\in \R$ is the weight of the loop at $i$th vertex and $w_{ij}\in\C$. Note that $r_i=0$ if $i$th vertex does not
have a loop. The weighted degree $d_i$ of a vertex $i\in V$ is given by $d_i =\sum_{j=1}^n |a_{ij}|.$ The Laplacian and the signless Laplacian matrices are
defined by \be L(G) = \mbox{diag}(\{d_i\}_{i=1}^n) - A(G) \,\, \mbox{and} \,\, Q(G) = \mbox{diag}(\{d_i\}_{i=1}^n) + A(G), \ee respectively \cite{graphs,BaKaPa,CveRowSim07}. Notice that
loops, even though apparent in the
adjacency matrix $A(G)$, do not appear in the Laplacian matrix $L(G).$ The above constructions of $L(G)$ and $Q(G)$ will lead to diagonally dominant matrices, i.e., a matrix $M$ where $|M_{i,i}| \ge \sum_{j \neq i}|M_{i,j}|$. 

For a weighted edge from vertex $i$ to $j$ with weight $w_{ij}\in\C,$ we assume $w_{ij}=r_{ij}e^{i\theta_{ij}}, r_{ij}>0, 0\leq \theta \leq \pi.$ We consider $\overline{w_{ij}}= r_{ij} e^{-i\theta_{ij}} -\pi\leq \theta_{ij}\leq 0, \sqrt{w_{ij}}=\sqrt{r_{ij}}e^{i\theta_{ij}/2}, \sqrt{\overline{w_{ij}}}=\sqrt{r_{ij}}e^{-i\theta_{ij}/2}.$ Thus, $(\sqrt{w_{ij}})^2=w_{ij}$ and $\sqrt{w_{ij}} \sqrt{\overline{w_{ij}}}=r_{ij}.$

\begin{lemma}\label{Lem:cond}
Let $G=(V,E)$ be a weighted directed graph without loops. Then $L(G)$ and $Q(G)$ are Hermitian and positive semidefinite matrices.
\end{lemma}
\pf Assume that $w_{ij}\in \C$ is the weight of an edge $(i,j)$. Then define $$(M^{-})_{v,e} =\left\{
  \begin{array}{ll}
  \sqrt{w}_{ij}, & \hbox{if $w_{ij}\in\C\setminus\R$ and $v$ is initial vertex of nonloop edge $e$} \\
 -\sqrt{\overline{w}}_{ij}, & \hbox{if $w_{ij}\in\C\setminus\R$ and $v$ is terminal vertex of nonloop edge $e$} \\
 \sqrt{|w_{ij}|}, & \hbox{if $0 > w_{ij}\in \R$ and $v$ is initial or terminal vertex of nonloop edge $e$} \\
  -\sqrt{|w_{ij}|}, & \hbox{if $0 < w_{ij}\in \R$ and $v$ is initial vertex of nonloop edge $e$} \\
  \sqrt{|w_{ij}|}, & \hbox{if $0 < w_{ij}\in \R$ and $v$ is terminal vertex of nonloop edge $e$} \\
0, & \hbox{otherwise}
  \end{array}
 \right. $$ and $$(M^{+})_{v,e} =\left\{
  \begin{array}{ll}
  \sqrt{w}_{ij}, & \hbox{if $w_{ij}\in\C\setminus\R$ and $v$ is initial vertex of nonloop edge $e$} \\
 \sqrt{\overline{w}}_{ij}, & \hbox{if $w_{ij}\in\C\setminus\R$ and $v$ is terminal vertex of nonloop edge $e$} \\
 \sqrt{|w_{ij}|}, & \hbox{if $0 < w_{ij}\in \R$ and $v$ is initial or terminal vertex of nonloop edge $e$} \\
  -\sqrt{|w_{ij}|}, & \hbox{if $0 > w_{ij}\in \R$ and $v$ is initial vertex of nonloop edge $e$} \\
  \sqrt{|w_{ij}|}, & \hbox{if $0 > w_{ij}\in \R$ and $v$ is terminal vertex of nonloop edge $e$} \\
0, & \hbox{otherwise}
  \end{array}
 \right. $$ for any $v\in V$ and $e\in E.$ Then it is easy to verify that $L(G)=M^{-}(M^{-})^\dagger$ and $Q(G)=M^{+}(M^{+})^\dagger$ where $^\dagger$ denotes conjugate transpose. Therefore, the result follows.


 From the construction of $M^+$ and $M^-$ in Lemma \ref{Lem:cond}, it follows that \begin{align}\label{cond:zeroevL} x^\dagger L(G)x &=& \sum_{i\neq j, (i,
 j)\in E, 0< w_{ij}\in\R} |w_{ij}||x_i -x_j|^2 + \sum_{i\neq j, (i,
 j)\in E, 0> w_{ij}\in\R} |w_{ij}||x_i +x_j|^2 \notag \\ && + \sum_{i\neq j, (i,
 j)\in E, w_{ij}\in\C\setminus\R} |\sqrt{\overline{w_{ij}}}x_i - \sqrt{w_{ij}} x_j|^2\end{align}

 and \begin{align}\label{cond:zeroevSL} x^\dagger Q(G)x &=& \sum_{i\neq j, (i,
 j)\in E, 0< w_{ij}\in\R} |w_{ij}||x_i +x_j|^2 + \sum_{i\neq j, (i,
 j)\in E, 0> w_{ij}\in\R} |w_{ij}||x_i -x_j|^2 \notag \\ && + \sum_{i\neq j, (i,
 j)\in E, w_{ij}\in\C\setminus\R} |\sqrt{\overline{w_{ij}}}x_i + \sqrt{w_{ij}} x_j|^2\end{align}



In the following theorem we provide a necessary and sufficient condition for a connected loopless weighted digraph having a signless Laplacian eigenvalue zero. Recall that, a digraph
is said to be connected if it is connected without considering the directions of the edges.

\begin{theorem} \label{thm:zeroeigL0}
Let $G$ be a connected weighted digraph without selfloops. The least eigenvalue of the signless Laplacian matrix of a loopless connected weighted digraph is $0$ if and only if $$(-1)^{p_+ + |P|}\prod_{(i,j)\in P, w_{ij}\in\C\setminus\R}\frac{w_{ij}}{|w_{ij}|} = (-1)^{p'_+ + |P'|}\prod_{(ij)\in P', w_{ij}\in\C\setminus\R}\frac{w_{ij}}{|w_{ij}|}$$ holds for any two directed paths $P$ and $P'$ with numbers of links having positive real weights are $p_+$ and $p'_+$ respectively between any fixed two vertices and $|P|$ denotes the number of edges having nonreal weights in $P$.
\end{theorem}
\pf Assume that the least eigenvalue of the signless Laplacian $Q(G)$ of $G$ has an eigenvalue zero, that is, $x^HQ(G)x=0$ for some non-zero vector $x.$ From (\ref{cond:zeroevSL}), it is obvious that for such $x,$ $x_i=-x_j$ if $0< w_{ij}\in\R;$ $x_i=x_j$ if $0> w_{ij}\in\R;$ $x_i=-\frac{w_{ij}}{|w_{ij}|}x_j$ if $ w_{ij}\in\C\setminus\R.$ Let $P\equiv (u=i_1, i_2, \hdots, i_{k_1}=v)$ and $P'\equiv (u=i'_1, i'_2,i'_3, \hdots, i'_{k_2}=v)$ be two distinct directed paths from the vertex $u$ to the vertex $v.$ Then, for the path $P,$ $$x_u=(-1)^{p_+ + |P|}\prod_{(i,j)\in P, w_{ij}\in\C\setminus\R}\frac{w_{ij}}{|w_{ij}|} \,\, x_v;$$ and for the path $P',$ $$x_u=(-1)^{p'_+ + |P'|}\prod_{(ij)\in P', w_{ij}\in\C\setminus\R}\frac{w_{ij}}{|w_{ij}|} \,\, x_v.$$ Further, $x_u\neq 0$ and $x_v\neq 0$ since otherwise $x=0$ follows from (\ref{cond:zeroevSL}) as the graph is connected. Hence the desired result follows.

Conversely, if the given condition is true for any two different directed paths for any pair of vertices in $G,$ a vector $x$ defined by $x_i=-x_j$ if $0< w_{ij}\in\R;$ $x_i=x_j$ if $0> w_{ij}\in\R;$ $x_i=-\frac{w_{ij}}{|w_{ij}|}x_j$ if $ w_{ij}\in\C\setminus\R$ will satisfy $x^\dagger Q(G)x=0.$ Hence the proof.

\begin{corollary} \label{thm:zeroeig}
The least eigenvalue of the signless Laplacian of a loopless connected weighted digraph having complex unit weights is equal to
$0$ if and only if
$$W(P) = (-1)^{p'-p}W(P')$$ holds for any two directed walks $P, P'$ of lengths $p$ and $p'$ respectively between any fixed two vertices where $W(P) $ (resp. $W(P')$) is the product
of the weights of the edges of $P$ (resp. $P'$). In particular, $0$ is a simple eigenvalue.
\end{corollary}

%

\begin{corollary}\label{cor:concomp2}
Let $G$ be a weighted digraph without loops with $n (>2)$ vertices. Assume that $0$ is a signless Laplacian eigenvalue of $G.$ Then the multiplicity of $0$ as a signless Laplacian
eigenvalue of $G$ is $k$ if and only if the graph is disconnected with $k$ connected components.
\end{corollary}

It is shown in \cite{CveRowSim07} that for a unweighted undirected connected graph $G$ without loops, the least eigenvalue of the signless Laplacian of $G$ is equal to
$0$ if and only if the graph is bipartite and $0$ is a simple eigenvalue. We mention that the condition provided in Theorem \ref{thm:zeroeig} for existence of zero eigenvalue of
weighted directed graph is a generalized version of the condition obtained for unweighted undirected graph in \cite{CveRowSim07}. That is, the condition in Theorem \ref{thm:zeroeig}
is satisfied for an unweighted undirected connected graph if and only if the graph is bipartite.

%

The following corollary provides a necessary and sufficient condition for existence of zero Laplacian eigenvalue of a weighted connected digraph.

\begin{corollary} \label{thm:zeroeigL00}
The least eigenvalue of the combinatorial Laplacian matrix of a loopless connected weighted digraph is equal to
$0$ if and only if
$$(-1)^{p_-}\prod_{(i,j)\in P, w_{ij}\in\C\setminus\R}\frac{w_{ij}}{|w_{ij}|} = (-1)^{p'_-}\prod_{(ij)\in P', w_{ij}\in\C\setminus\R}\frac{w_{ij}}{|w_{ij}|}$$ holds for any two directed walks $P, P'$ with number of links with negative real weights ares $p_-$ and $p'_-$ respectively between any fixed two vertices.
\end{corollary}
\pf The proof is similar to the proof of Theorem \ref{thm:zeroeigL0}.

\begin{corollary}
The least eigenvalue of the cominatorial Laplacian of a loopless connected weighted digraph having complex unit weights is equal to
$0$ if and only if $$W(P) = W(P')$$ holds for any two walks $P, P'$ between any two vertices where $W(P) $ (resp. $W(P')$) is the product
of the weights of the edges of $P$ (resp. $P'$). In particular, $0$ is a simple eigenvalue.
\end{corollary}

An alternative proof of the above corollary also can be found in \cite{BaKaPa}.

\begin{corollary}\label{cor:concomp1}
Let $G$ be a weighted digraph without loops with $n (>2)$ vertices. Assume that $0$ is a combinatorial Laplacian eigenvalue of $G.$ Then the multiplicity of $0$ as a combinatorial
Laplacian eigenvalue of $G$ is $k$ if and only if the graph is disconnected with $k$ connected components.
\end{corollary}


 As we mentioned above, loops with nonnegative weights have no effect on the combinatorial Laplacian matrix. Thus, we only consider signless Laplacian matrix when an weighted digraph contains at least one loop with positive real weight. It is easy to verify that, given an weighted digraph $G$ with at least one loop having positive real weight, we have \be\label{cond2} x^\dagger Q(G)x=x^\dagger Q(\widehat{G})x + \sum_{(i,i)\in E} r_i|x_i|^2\ee where $\widehat{G}$ is the subgraph of $G$ without considering loops. This also shows that $Q(G)$ is Hermitian and positive semidefinite.

\begin{lemma}\label{re:withloop}
$0$ can never be a signless Laplacian eigenvalue of a connected weighted digraph $G$ with loops (at least one vertex contains a loop) having positive weights.
\end{lemma}
\noin\pf Consider a connected weighted digraph $G$ with loops (at least one vertex contains a loop) having positive weights. If $0$ is a signless Laplacian eigenvalue of $G,$ from
(\ref{cond2}) we know that, there exists an $0\neq x\in\C^n$ such that $$x^\dagger Q(G)x=x^\dagger Q(\widehat{G})x + \sum_{(i,i)\in E} r_i|x_i|^2=0$$ where $\widehat{G}$ is the subgraph of $G$ without loops. Assume that the $k$-th vertex contains the loop. For $x^\dagger Q_Gx$ to be zero, $x_k$ has to be zero since $r_k$ is positive. Further, since the graph is connected, $k$th vertex is
linked with $m$ (say) vertices $k_1, k_2, \hdots, k_m$ for some $m$ which implies
$x_{k_j}=0$ for $j=1,\hdots, m$ which further implies that $x_j=0$ for all $j=1, \hdots, n$ since $k_j$ vertices are linked with other vertices and $x^\dagger Q(\widehat{G})x=0$
for all $(i,j)\in E.$


\subsection{Weighted digraphs with loops having at least one loop with negative weight}
Recall the definitions of Laplacian and signless Laplacian matrices associated with a weighted digraph. Observe that loops, even though apparent in the
adjacency matrix $A(G)$ do not reflect in Laplacian matrix when the loops are having positive weights, and in signless Laplacian matrices when the loops are having negative weights.
Thus, for weighted digraphs with both nonnegative and negative weighted loops (at least one of the loops has negative weight), we introduce a new matrix, which we call \emph{signed
Laplacian}, denoted by $L_-(G)$ and $L_{\pm}(G)$ when $G$ has all the loops with nonpositive weights, and when $G$ contains loops with positive weights as well as negative weights,
respectively by \be\label{def:signedL} L_-(G) = R_-(G) + Q(\widehat{G}) \,\, \mbox{and} \,\, L_{\pm}(G) = R_{\pm}(G) + Q(\widehat{G})\ee where $R_-(G)=R_{\pm}(G)=\diag\{r_1, r_2,
\hdots, r_n\},$ $r_j\in\R$ denotes the weight of the loop at $j$th vertex, $j=1,\hdots, n$ and $Q(\widehat{G})$ is the signless Laplacian matrix of the graph $\widehat{G}$ constructed from $G$ without considering the loops.
Obviously, $L_-(G)$ and $L_{\pm}(G)$ are Hermitian matrices. In order to simplify the notation we denote $R(G)=R_-(G)=R_{\pm}(G).$

\begin{lemma}\label{Lem:nsL}
Given a weighted digraph $G$ with nonpositive loops, $L_-(G)$ is semi-definite if
$$\max_i |r_i| \leq \lam_{\min}Q(\widehat{G})$$ where $r_i\leq 0, i=1:n$ are the weights of the
loops present in the graph and $\lam_{min}Q(\widehat{G})$ denotes the minimum eigenvalue of $Q(\widehat{G}).$
\end{lemma}
\noin\pf For any unit vector $0\neq x\in\C^n,$ we have \be\label{pf:psd}-x^{\dagger}R(G)x \leq \max_i|r_i| \,\, \mbox{and} \,\,
x^{\dagger}Q(\widehat{G})x \geq
\lam_{\min}(Q(\widehat{G})).\ee In order to show that $L_-(G)$ is positive semi-definite, for any non-zero unit vector $x\in\C^n,$ we must have
$x^{\dagger}L_-(G)x=x^{\dagger}R(G)x+x^{\dagger}Q(\widehat{G})x \geq 0.$ If the given condition is satisfied, the proof follows from (\ref{pf:psd}).

Given a weighted digraph, the lemma declares that if the maximum of the modulus of weights of the loops do not exceed the minimum signless Laplacian eigenvalue of the graph
without considering the loops, then the signed Laplacian corresponding to the given graph will be positive semi-definite.


\begin{example}
\begin{enumerate} \item
Consider the graph given in Figure \ref{Fig:nwt}. The signed Laplacian matrix with negative loops associated with $G$ is given by
\beano L_-(G) &=& \bmatrix{1 &1 & 1 & -1\\ 1 & 1 & 1 & -1\\ 1 & 1 & 1 & -1 \\ -1 & -1 & -1 & 1}\\ &=& \bmatrix{-2 & 0 & 0 & 0 \\ 0 & -2 & 0 & 0\\ 0 & 0 & -2 & 0\\ 0 & 0 & 0& -2} +
\bmatrix{3 &1 & 1 & -1\\ 1 & 3 & 1 & -1\\ 1 & 1 & 3 & -1 \\ -1 & -1 & -1 & 3} \\ &=& R(G) + Q(\widehat{G}).\eeano Note that, eigenvalues of $Q(\widehat{G})$ are $2, 2, 2, 6.$

\begin{figure}[h!]$$\xymatrix{\ar@(ul,dl)_{-2}1\ar@{-}[r]^{\mathrm{1}} \ar@{-}[d]_{\mathrm{1}} \ar@{-}[rd]^{\mathrm{1}} & 2\ar@(ur,dr)^{-2}\ar@{-}[d]^{\mathrm{-1}} \\
\ar@(dl,ul)^{-2}3\ar@{-}[r]_{\mathrm{-1}}\ar@{-}[ru]_{\mathrm{-1}} & \ar@(dr,ur)_{-2}4}$$\caption{Graph with negative weighted loops}\label{Fig:nwt}\end{figure}

\item Consider the weighted digraph given in Figure \ref{Fig:nwt2} \begin{figure}[h!]$$\xymatrix{\ar@(ul,dl)_{-0.2}1\ar[r]^{\mathrm{i}} \ar[d]_{\mathrm{i}} & 2\ar@(ur,dr)^{-0.2}\\
    \ar@(dl,ul)^{-0.2}3\ar[ru]_{\mathrm{i}}}$$\caption{Graph with negative weighted loops}\label{Fig:nwt2}\end{figure} The eigenvalues of $L_-(G)$ are given by $0.0679,
    1.8000, 3.5321.$

\end{enumerate}
\end{example}

\begin{proposition}\label{Proposition:nsL}
Let $G$ be a weighted digraph with loops having nonpositive weights. If the subgraph $\widehat{G}$ obtained from $G$ by removing the loops has signless Laplacian eigenvalue zero then
$L_-(G)$ is not positive semi-definite.
\end{proposition}
\noin\pf Let $x$ be the unit eigenvector corresponding to the zero Laplacian eigenvalue of $\widehat{G}.$ Then $x^{\dagger}L_-(G)x=x^{\dagger}R(G)x <0$ since $R$ contains at least one
loop with negative weight.

\begin{theorem}\label{thm:negwtloop}
Let $G$ be a connected weighted digraph with nonpositive loop weights and $n$ number of vertices. Let $\lam$ be a signless Laplacian eigenvalue of $\widehat{G}$ of algebraic
multiplicity $k.$ Then the number of zero eigenvalues of $L_-(G)$ is $k$ if $|r_i|=\lam$ for all $i\in V(G).$
\end{theorem}
\noin\pf Consider $L_-(G)=R(G)+Q(\widehat{G}).$ We know that $0$ is an eigenvalue of $L_-(G)$ if an only if there exists a unit vector $x\in\C^n$ such that $x^{\dagger}L_-(G)x=0,$
which implies $x^{\dagger}R(G)x=x^{\dagger}Q(\widehat{G})x.$ Assume that $\lam$ is an eigenvalue of $Q(\widehat{G})$ with algebraic multiplicity $k.$ Since $Q(\widehat{G})$ is
Hermitian, there exists unit orthogonal vectors $x_1, x_2, \hdots, x_k$ such that $x_i^{\dagger}Q(\widehat{G})x_i=\lam$, $i=1, \hdots, k.$ Set $R(G)=-\lam I_n,$ where $I_n$ is the
identity matrix of order $n.$ Then $x_i^{\dagger}L_-(G)x_i=0$ for all $i.$ Since $x_i$s are orthonormal vectors, algebraic multiplicity of eigenvalue $0$ of $L_-(G)$ is $k.$

\begin{remark}
Note that, a connected weighted digraph with all negative weighted loops can have zero signed Laplacian eigenvalues of multiplicity more than one. In particular, consider any
connected weighted complete digraph $G$ where all the loops are present having weights equal to one of the eigenvalues of $Q(\widehat{G}).$ For example, consider Figure
\ref{Fig:nwt}.
\end{remark}

Now we consider weighted digraphs with loops having both positive and negative weights. Then we have the following lemma.

\begin{lemma}\label{Lem:pnL}
Let $G$ be a weighted digraph with loops having at least one negative and at least one positive weighted loops. Assume that $G$ has $n$ vertices, $k$ are having positive
loop weights, $l$ are having negative loop weights such that $k+l\leq n.$ Let $r_1^+, \hdots, r_k^+$ be the positive weights and $r_{k+1}^-, \hdots, r_l^-$ the negative
weights such that $r_j^{\pm}=r_j-d_j,$ for some $r_j\in\R,$ $j=1,\hdots,k,k+1,\hdots,l,\hdots, n$ where $d_j$ is the degree of $j$th vertex of $\widehat{G}$. Then $L_{\pm}(G)$ is
positive semi-definite if and only if $$\diag\{r_1, r_2, \hdots, r_n\} + A(\widehat{G})$$ is positive semi-definite.
\end{lemma}
\noin\pf Without loss of generality, assume that the first $k$ vertices are having loops with positive weights, second $l$ vertices are having loops with negative weights,
and the remaining (if any) vertices have no loops. For any nonzero $x\in\C^n,$ \beano x^{\dagger}L_{\pm}(G)x &=& x^{\dagger}\bmatrix{R_+ & 0 & 0 \\ 0 & R_- &  0 \\ 0 & 0 & R_0}x
+ x^{\dagger}Q(\widehat{G})x \\ &=& x^{\dagger}(\diag\{r_1, r_2, \hdots, r_n\} + A(\widehat{G}))x\eeano where $R_+=\diag\{r_1^+,\hdots,r_k^+\},$ $R_-=\diag\{r_1^-,\hdots, r_l^-\}$ and
$R_0=\diag\{r_{l+1}, \hdots, r_n\}.$ Hence, the result follows.


\begin{theorem}
Let $G$ be a weighted digraph with loops having both positive and negative weighted loops. The number of zero eigenvalues of $L_{\pm}(G)$ is equal to the number of zero eigenvalues of
$\diag\{r_1, r_2, \hdots, r_n\} + A(\widehat{G})$ where $r_j^{\pm}=r_j-d_j, j=1,\hdots, n$ and $r_j^{\pm}$ is the weight of the loop at $j$th vertex.
\end{theorem}
\noin\pf The proof follows by Lemma \ref{Lem:pnL}.

\begin{example}
Consider the weighted digraph in Figure \ref{Fig:pmwt}. \begin{figure}[h!]$$\xymatrix{ \ar@(ul,dl)_{-1}1\ar@/^/[rrr]^{-2}& 2& 3  & \ar@(ur,dr)^{2}4\ar@/_/[lll]}$$\caption{Graph $G$
with negative and positive weighted loops}\label{Fig:pmwt}\end{figure}
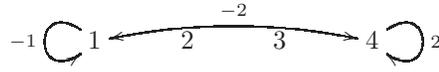 The signed Laplacian is $$L_{\pm}(G)=\bmatrix{1 & 0 & 0 & -2 \\ 0 & 0 & 0 & 0\\ 0 & 0 & 0 & 0\\ -2 & 0 & 0 &
4}=\bmatrix{-1 & 0 & 0 & 0 \\ 0 & 0 & 0 & 0\\ 0 & 0 & 0 & 0\\ 0 & 0 & 0 & 2} + \bmatrix{2 & 0 & 0 & -2 \\ 0 & 0 & 0 & 0\\ 0 & 0 & 0 & 0\\ -2 & 0 & 0 & 2}.$$

\end{example}

\section{Graph structure for pure and mixed states}
The approach of identifying the density matrix representations of quantum states by density matrices of unweighted undirected
graphs was introduced in \cite{BrGhSe06}  and extended in \cite{HasJoa} for weighted graphs. However, the recent development of
signless Laplacian matrix associated with a graph has not been gainfully used in both \cite{BrGhSe06} and \cite{HasJoa} to define density matrix associated with the graph. Thus, the results in
\cite{HasJoa} could not capture interesting connections between the properties of density matrices defined by a graph and topology of the graph. In this section, we use the Laplacian
matrices defined in the last section to define density matrix associated with a weighted digraph. Further, we describe how the
topological structure of the graph dictates whether the
corresponding density matrices represent pure or mixed states.

\subsection{Pure States and Mixed States}
 Recall that density matrix representation of a quantum state is a Hermitian positive semi-definite
matrix with unit trace. The density matrix $\sigma$ corresponding to a state is called pure if $\tr(\sigma^2)=1$ and mixed if $\tr(\sigma^2)<1.$ A quantum state, in general, can also
be represented as \be \sigma = \sum_{i} p_i |\psi_i\rangle\langle\psi_i|,\ee
where $0\neq |\psi_i\rangle\in\C^2$ with norm one and $\sum_i p_i =1, 0\leq p_i \leq 1$. Thus $\sigma$ is a convex combination of rank one matrices, in particular, rank one
projections. If $\sigma$ is just a projection with rank one then $\sigma$ is a pure state, otherwise, a mixed state.


We define density matrices associated with a weighted digraph. We denote an weighted complete bipartite digraph without loops of order $n$ by $K_n.$

\begin{definition} The density matrix $\sigma_G$ associated with an weighted digraph $G$ are given by
\be \sigma_G := \frac{1}{\tr(K(G))} K(G) \ee where \begin{itemize} \item $K(G)=L(G)$ when $G$ is without a loop \item $K(G)=Q(G)$ when either $G$ is without a loop or having loops
with nonnegative weights \item $K(G)=L_-(G)$ when $G$ is with loops having nonpositive weights and $L_-(G)$ is positive semi-definite \item $K(G)=L_{\pm}(G)$ when $G$ contains loops
with positive weights, negative weights and $L_{\pm}(G)$ is positive semi-definite.\end{itemize} \end{definition}

\begin{theorem}\label{thm:pures} The density matrix defined by Laplacian or signless Laplacian matrix of a weighted digraph $G$ without loops has rank one if and only if the graph is
$K_
2$ or $\widehat{K}_2 := K_2 \sqcup  v_1 \sqcup v_2 \sqcup \hdots v_{n-2}.$, where $v_1$, $v_2$, $\hdots$, $v_{n-2}$ are isolated vertices.\end{theorem}

\noin\pf Assume that $\sigma_G$ has rank one and $G$ contains $n$ vertices. Then $\sigma_G$ has eigenvalue $1$ with multiplicity one (since trace of $\sigma_G =1$) and $0$ is an
eigenvalue of multiplicity $n-1.$ If $n=2$ then obviously $G=K_2.$ If $n\neq 2$ then by Corollary \ref{cor:concomp1}, $G$ contains $n-1$ connected
components. Thus $G = \widehat{K}_2.$

Conversely, suppose $G = K_2$ or $\widehat{K}_2.$ Then the eigenvalues of $\sigma_G$ are $0$ with multiplicity $n-1$ for $G=\widehat{K}_2$ and multiplicity $1$ for $G=K_2$, and $1$
with multiplicity one. Hence the result follows.

\begin{remark}
We mention that the same result has been obtained in \cite{BrGhSe06} for unweighted undirected graphs.
\end{remark}

\begin{corollary}\label{thm:pure2}
Let $G$ be a weighted digraph without loops isomorphic to $K_2$ or $\widehat{K}_2.$ Then $\sigma_G$ constructed by $L(G)$ or $Q(G)$ represents a pure state.

\end{corollary}

\noin\pf The density matrix $\sigma_G$ has a simple eigenvalue $1$ and other eigenvalues are zeros. Since trace of any matrix is sum of the eigenvalues of the matrix, we
have $\tr(\sigma_G)=1$ and $\tr(\sigma_G^2) = 1.$ Thus the result follows.

\begin{corollary}\label{thm:mixed}
Let $G$ be a weighted digraph without loops of order $n$ that is not isomorphic to $K_2$ and $\widehat{K}_2.$ Then $\sigma_G$
constructed by $L(G)$ or $Q(G)$ represents a mixed state.
\end{corollary}

\noin\pf Let the eigenvalues of $\sigma(G)$ be $\lam_1\leq \lam_2\leq \hdots\leq \lam_n.$ By the definition of $\sigma_G$ we have $\tr(\sigma_G)=1$ as $\sum_{i=1}^n
\frac{\lam_i}{d(G)} =1$ where $d(G)=\sum_{i=1}^n \lam_i.$ Then the eigenvalues of $\sigma_G^2$ are $\frac{\lam_1^2}{d(G)^2}, \frac{\lam_2^2}{d(G)^2}, \hdots, \frac{\lam_n^2}{d(G)^2}.$
Thus \beano \tr(\sigma_G^2) &=& \frac{\sum_{i=1}^n \lam_i^2}{d(G)^2} = \frac{d(G)^2 - 2\sum_{i\neq j, i,j=1}^n \lam_i\lam_j}{d(G)^2} < 1.\eeano Hence $G$ represents a mixed state.

\begin{example}
\begin{enumerate}
\item Consider the graph in Figure \ref{Fig:wt1} which represents a pure state. 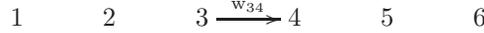
\begin{figure}[h!]$$\xymatrix{ 1& 2& 3 \ar[r]^{\mathrm{w_{34}}} & 4 & 5 & 6 }$$\caption{Pure state
    given by $K_2$ along with $4$ isolated nodes}\label{Fig:wt1}\end{figure}
\item Consider the graph in Figure \ref{Fig:wt2} which represents a mixed state. \begin{figure}[h!]$$\xymatrix{ 1& 2\ar[r]^{\mathrm{w_{23}}}& 3  & 4\ar[l]^{\mathrm{w_{43}}} & 5 &
    6 }$$\caption{Mixed state with $6$ vertices}\label{Fig:wt2}\end{figure}
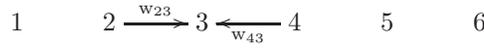

\end{enumerate}
\end{example}

The digraph $O_1$ denotes a digraph with one vertex and the vertex contains a directed weighted loop.

\begin{theorem}\label{thm:pureswithloop} The density matrix of order $n$ defined by signless Laplacian matrix associated with a weighted digraph $G$ with loops having nonnegative
weights has rank one if and only if the graph is $\widehat{O}_1 := O_1 \sqcup  v_1 \sqcup v_2 \sqcup \hdots v_{n-1}$, where $v_1$, $v_2$, $\hdots$, $v_{n-1}$ are isolated vertices
without loops.\end{theorem}

\noin\pf Assume that the density matrix $\sigma_G$ of $G$ constructed by the signless Laplacian of the graph and $G$ contains $n$ vertices with at least one vertex contains a directed
loop. Obviously, the matrix $\sigma_G$ has rank $ 1$ if and only if any submatrix of $\sigma_G$ of order $\geq 2$ is singular. Without loss of generality assume that the first vertex
$v_1$ is attached with a directed loop. Then the following cases arise.

Case-I: The vertex $v_1$ is linked by a directed edge with another vertex say the second vertex $v_2.$ In this case, if we consider the $2\times 2$ submatrix of $\sigma_G$ constructed
by the intersection of the 1st row, 2nd row, 1st column and 2nd column of $\sigma_G$ then this is a matrix with nonzero determinant. Hence rank of $\sigma_G$ is at least $2.$

Case-II: Any two vertices without loops say $v_i$ and $v_j$ are linked by a directed edge.  In this case, if we consider the $2\times 2$ submatrix of $\sigma_G$ constructed by the
intersection of the 1st row, $i$th row, 1st column and $i$th column of $\sigma_G$ then this is a matrix with nonzero determinant. Hence rank of $\sigma_G$ is at least $2.$

Case-III: All the vertices are isolated without loops except two vertices say the 1st and the 2nd vertex. Then, if we consider the $2\times 2$ submatrix of $\sigma_G$ constructed by
the intersection of the 1st row, 2nd row, 1st column and 2nd column of $\sigma_G$ then this is a matrix with nonzero determinant. Hence rank of $\sigma_G$ is at least $2.$

Case-IV: All the vertices are isolated without loops except the 1st vertex. Then the density matrix $\sigma_G$ is of rank $1$.

Hence the desired result follows.

\begin{corollary}\label{thm:pure2withloop}
Let $G$ be a weighted digraph with loops having nonnegative weights and is isomorphic to $\widehat{O}_1.$ Then $\sigma_G$ defined
by the signless Laplacian of $G$ represents a pure state.
\end{corollary}

\begin{corollary}\label{thm:mixedwithloop}
Let $G$ be a weighted digraph with loops having nonnegative weights of order $n$ that is not isomorphic to $\widehat{O}_1.$
Then $\sigma_G$ defined by the signless Laplacian of $G$ represents a mixed state.
\end{corollary}


\begin{example}
\begin{enumerate}
\item Consider the graph in Figure \ref{Fig:wwt1} which represents a pure state. 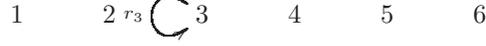
\begin{figure}[h!]$$\xymatrix{ 1& 2& 3\ar@(ul,dl)_{r_3} & 4 & 5 & 6 }$$\caption{Pure state given
    by $O_1$ along with a single loop}\label{Fig:wwt1}\end{figure}
\item Consider the graph in Figure \ref{Fig:wwt2} which represents a mixed state. 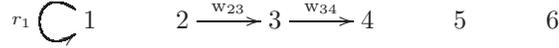
\begin{figure}[h!]$$\xymatrix{ \ar@(ul,dl)_{r_1}1& 2\ar[r]^{\mathrm{w_{23}}}& 3
    \ar[r]^{\mathrm{w_{34}}}& 4 & 5 & 6 }$$\caption{Mixed state with $6$ vertices}\label{Fig:wwt2}\end{figure}

\end{enumerate}
\end{example}

%
%
%
%
%

\begin{remark} Realization of $1$-qubit by weighted digraph: Consider the graph $G=K_2$ for a pure state with edge weight $w\in\S_1^+.$ Then the corresponding density matrix with
respect to the
Laplacian matrix is given by $$\sigma_G =\frac{1}{2} L(G) = \frac{1}{2}\bmatrix{1 & -w \\ - \overline{w} & 1},$$
$\mbox{where}~ w =e^{i\phi}, 0\leq \phi\leq 2\pi$. The eigenvalues of
$\sigma_G$ are $0$ and $1$ corresponding to eigenvectors $|\psi_1\rangle = \frac{1}{\sqrt{2} |z_1|}\bmatrix{z_1 \\ \overline{w}z_1}$
and
$|\psi_2\rangle = \frac{1}{\sqrt{2} |z_2|} \bmatrix{z_2 \\ -\overline{w}z_2}$ respectively, where $0\neq z_1, z_2\in\C$.
Thus the pure state is given by $\sigma = |\psi_2\rangle\langle\psi_2|.$ Setting
$z_2 = r e^{i\theta}, |z_2|=r>0, 0\leq \theta \leq 2\pi,$ the vector representation of the pure state is given by
\beano |\psi\rangle &=& e^{i\theta} (\frac{1}{\sqrt{2}} |0\rangle - \frac{1}{\sqrt{2}} e^{-i\phi}|1\rangle),
\\ &\equiv& \frac{1}{\sqrt{2}} |0\rangle - \frac{1}{\sqrt{2}} e^{-i\phi}|1\rangle\eeano where $|0\rangle=\bmatrix{1 \\ 0}$ and $|1\rangle=\bmatrix{0 \\ 1}.$

Further, the density matrix with respect to the signless Laplacian matrix is given by
$$\sigma_G =\frac{1}{2} Q(G) = \frac{1}{2}\bmatrix{1 & w \\ \overline{w} & 1}.$$ Following a similar approach, as above,
the corresponding vector representation of the pure state is given by
$$|\psi\rangle \equiv \frac{1}{\sqrt{2}} |0\rangle + \frac{1}{\sqrt{2}} e^{-i\phi}|1\rangle, 0\leq \phi\leq 2\pi. $$

\end{remark}


Now we consider weighted graphs with loops having nonpositive weights. We denote $S_n, n\geq 2$, a star graph with $n$ vertices.

\begin{theorem}
Consider a weighted digraph $G$ consisting of a weighted digraph $\widehat{G}$ without loops having $n$ number of vertices and loops at each vertex of $\widehat{G}$ with equal
weights $-\lam$ where $\lam$ is a signless Laplacian eigenvalue of $\widehat{G}$ with multiplicity $n-1.$ Then $\sigma(G)=\frac{1}{\tr(L_-(G))}L_-(G)$ represents a pure state.
\end{theorem}
\noin\pf By Lemma \ref{Lem:nsL}, $\sigma(G)$ is positive semi-definite. However, $n-1$ number of eigenvalues of $L_-(G)$ are zero since $\lam$ is an eigenvalue of $Q(\widehat{G})$ of
algebraic multiplicity $n-1.$ Therefore, rank of $\sigma(G)$ is one. Hence the result follows.

\begin{corollary}
Consider a weighted digraph $G$ consists of a weighted digraph $\widehat{G}$ without loops having $n$ number of vertices and loops at each vertex of $\widehat{G}$ with equal
weights $-\lam$ where $\lam$ is a signless Laplacian eigenvalue of $\widehat{G}$ with multiplicity $k< n-1.$ Then $\sigma(G)=\frac{1}{\tr(L_-(G))}L_-(G)$ represents a mixed state.

\end{corollary}

\begin{example} \begin{enumerate} \item Consider $G=K_n$ along with loops at each vertex of equal weights $-n/(n-1).$ Then
$\sigma(G)=\frac{1}{\tr(L_-(G))}L_-(G)$ represents a pure state. For instance, consider $n=3$ in Figure \ref{Fig:nwt3}.
\begin{figure}[h!]$$\xymatrix{\ar@(ul,dl)_{-1.5}1\ar@{-}[r]^{\mathrm{1}} \ar@{-}[d]_{\mathrm{1}} & 2\ar@(ur,dr)^{-1.5}\\ \ar@(dl,ul)^{-1.5}3\ar@{-}[ru]_{\mathrm{1}}}$$\caption{Pure
state given by $K_3$ along with negative weighted loops}\label{Fig:nwt3}\end{figure}
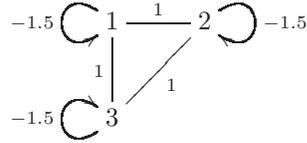
\item Consider $G=S_n$ along with loops at each vertex of equal weights $-1.$ Then $\sigma(G)=\frac{1}{\tr(L_-(G))}L_-(G)$ represents a mixed state. For
    example, consider $n=4$ in Figure \ref{Fig:nwt4}. \begin{figure}[h!]$$\xymatrix{\ar@(ul,dl)_{-1}1 \ar@{-}[d]^{\mathrm{1}} \\ \ar@(ul,dl)_{-1}2\ar@{-}[r]^{\mathrm{1}}
    \ar@{-}[d]_{\mathrm{1}} & 3\ar@(ur,dr)^{-1}\\ \ar@(dl,ul)^{-1}4}$$\caption{Mixed state given by $S_4$ along with negative weighted loops}\label{Fig:nwt4}\end{figure}
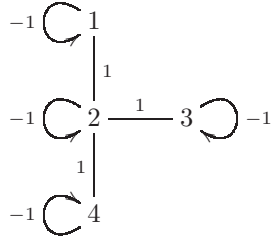
\end{enumerate}
\end{example}

Now we consider graphs with both positive and negative weighted loops.

\begin{theorem}\label{Thm:pnL}
Let $G$ be a weighted digraph with loops having at least one negative and at least one positive weighted loops. Assume that $G$ has $n$ vertices, $k$ are having positive
loop weights, $l$ are having negative loop weights such that $k+l\leq n.$ Let $r_1^+, \hdots, r_k^+$ be the positive weights and $r_{k+1}^-, \hdots, r_l^-$ the negative
weights such that $r_j^{\pm}=r_j-d_j, j=1,\hdots,k,k+1,\hdots,l,\hdots, n$ where $d_j$ is the degree of $j$th vertex of $\widehat{G}$. Then $G$ represents a pure state if and only if
$$\diag\{r_1, r_2, \hdots, r_n\} + A(\widehat{G})$$ has rank one.
\end{theorem}
\noin\pf The proof follows from the construction of $L_{\pm}(G)$ and Lemma \ref{Lem:pnL}.

However, we can construct a class of pure states by using the construction mentioned in the following corollary.
\begin{corollary}
 Consider a digraph $\widehat{G}$ without loops having $n$ vertices that represents a pure state obtained by signless Laplacian matrix $Q(\widehat{G})$, that is, only two vertices of
 $G$, say $i$th and $j$th of $\widehat{G}=(V,E)$ are linked having edge weight $w_{ij}\in\C$, rest of the vertices are isolated. Define $r_i^+=r_i^2-|w_{ij}|$ and
 $r_j^-=r_j^2-|w_{ij}|$ where $r_i, r_j\in\R_+$ such that $r_i^2 + r_j^2=1$ and $r_ir_j=|w_{ij}|.$ Then the graph $G$ constructed by $\widehat{G}$ along with loops introduced at the
 $i$th and $j$th vertices having weights $r_i^+$ and $r_j^-$ respectively, provides a pure state defined by $L_{\pm}(G).$
\end{corollary}
\noin\pf Note that, all the entries of $L_{\pm}(G)=(l_{pq})$ are given by $$(L_{\pm}(G))_{pq}=\left\{
                                                                                                          \begin{array}{ll}
                                                                                                            r_i^2, & \hbox{if $p=q=i$} \\
                                                                                                            r_j^2, & \hbox{if $p=q=j$} \\
                                                                                                            w_{ij}, & \hbox{if $p=j, q=j$} \\
                                                                                                            \overline{w}_{ij}, & \hbox{if $p=j, q=i$}\\
0 & \hbox{otherwise.}
                                                                                                          \end{array}
                                                                                                        \right.
$$ Obviously, $L_{\pm}(G)$ is Hermitian, positive semi-definite and $\tr(L_{\pm}(G))=1.$ Further, rank of $L_{\pm}(G)=1.$ Therefore, $G$ represents a pure state.

\section{Graph structure of entangled pure states}
In this section, we provide weighted digraphs whose density matrices represent entangled pure states. Because of the potential applications offered by pure entangled states, they are
of immense importance in quantum information and computation. This forms the basis for studying the properties of such states
from a graph theoretic approach.


%

\begin{enumerate}
\item \textit {Bell States}: For two-qubit systems, Bell states \cite{Einstein} are maximally entangled states represented as
\begin{eqnarray}
\label{Bell}
\left| \phi\right\rangle^{\pm}_{12}  &= &\frac{1}{\sqrt{2}} \left[ \,
\left| 00 \right\rangle_{12} \pm \left| 11 \right\rangle_{12} \,
\right],   \nonumber \\
\left| \psi\right\rangle^{\pm}_{12}  &=& \frac{1}{\sqrt{2}} \left[ \,
\left| 01 \right\rangle_{12} \pm \left| 10 \right\rangle_{12} \,
\right]\ .
\end{eqnarray}
For example, consider the graphs with four vertices in Figure \ref{Fig:bells}. 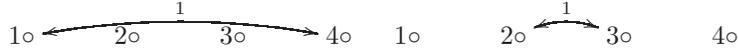
\begin{figure}[h!]$$\xymatrix{1\circ\ar@/^/[rrr]^{\mathrm{1}} & 2\circ & 3\circ &
\ar@/_/[lll]4\circ} \,\,\,\,\,\, \xymatrix{1\circ  & 2\circ\ar@/^/[r]^{\mathrm{1}} & 3\circ\ar@/_/[l] & 4\circ}$$\caption{$G_1$ and $G_2$}\label{Fig:bells}\end{figure} The density
matrices are given by $\sigma(G_i)=\frac{K(G_i)}{\tr(K(G_i))}$; $K(G_i)\in\{L(G_i), Q(G_i)\}$ where $\sigma(G_1)=
\left| \phi  \right\rangle _{12}^{+}  \left\langle \phi  \right|_{12}^{+}$ and $\sigma(G_2) = \left| \psi  \right\rangle _{12}^{+}  \left\langle \psi  \right|_{12}^{+}$. In order
to produce the Bell states of the form   $\frac{1}{\sqrt{2}}\left[|00\rangle +e^{i\delta} |11\rangle\right] $ and $\frac{1}{\sqrt{2}}\left[|01\rangle +e^{i\delta}
|10\rangle\right]$ using $G_1$ and $G_2$, one has to replace the edge weights by a factor $e^{i\delta}$ and edge will be unidirectional.

\item \textit {General $2$-qubit and $3$-qubit entangled states}: Consider the graph in Figure \ref{Fig:pmges}\begin{figure}[h!]$$\xymatrix{
    \ar@(ul,dl)_{|a|^2-|ab|}1\circ\ar@/^/[rrr]^{a\overline{b}}& 2\circ & 3\circ  & \ar@(ur,dr)^{|b|^2-|ab|}4\circ}$$\caption{Graph $G$}\label{Fig:pmges}\end{figure} for the
    general two-qubit state $\left|\Psi\right\rangle = a\left|00\right\rangle + b \left|11\right\rangle$, where $a,b\in\C\setminus\{0\}$ and $|a|^2+|b|^2=1.$ The density matrix
    associated with $G$ is given by $\sigma(G)=\frac{L_{\pm}(G)}{\tr(L_{\pm}(G))}=\left| \Psi  \right\rangle \left\langle \Psi  \right|$. The graph for a general $3$-qubit state,
    $\left|\Phi\right\rangle = a\left|000\right\rangle + b \left|111\right\rangle$, will follow similarly by considering $8$ vertices where only the first and the last vertices
    will be linked.

\item \textit {Three-qubit GHZ and W States}: Three qubit states can be separated into two inequivalent classes, namely GHZ class
and W class \cite{Greenberger,Dur}. These classes have distinct properties and cannot be converted into one another by performing
Stochastic Local Operations and Classical Communication (SLOCC). We have already shown that the graph of a general three-qubit GHZ state will be similar to Fig. (11). For specific cases of $a=b=\frac{1}{\sqrt{2}}$, the eight orthogonal GHZ states are,
\begin{eqnarray}
\label{GHZ}
\left| \psi\right\rangle^{(1),(2)}_{123}  &= &\frac{1}{\sqrt{2}} \left[ \,
\left| 000 \right\rangle\pm \left| 111 \right\rangle \,
\right],   \nonumber \\
\left| \psi\right\rangle^{(3),(4)}_{123}  &= &\frac{1}{\sqrt{2}} \left[ \,
\left| 001 \right\rangle \pm \left| 110 \right\rangle \,
\right],   \nonumber \\
\left| \psi\right\rangle^{(5),(6)}_{123}  &= &\frac{1}{\sqrt{2}} \left[ \,
\left| 010 \right\rangle \pm \left| 101 \right\rangle \,
\right],   \nonumber \\
\left| \psi\right\rangle^{(3),(4)}_{123}  &= &\frac{1}{\sqrt{2}} \left[ \,
\left| 011 \right\rangle \pm \left| 100 \right\rangle \,
\right],   \nonumber \\
\end{eqnarray}
The graphs corresponding to Eq. (12) are given in Figures \ref{Fig:GHZ3q1},\ref{Fig:GHZ3q2}, \ref{Fig:GHZ3q3}, \ref{Fig:GHZ3q4}.
\begin{figure}[h!]$$\xymatrix{1\circ\ar@/^/[rrrrrrr]^{\mathrm{1}}  & 2\circ & 3\circ & 4\circ & 5\circ & 6\circ & 7\circ &
8\circ\ar@/_/[lllllll]}$$\caption{$G_1$}\label{Fig:GHZ3q1}\end{figure}

\begin{figure}[h!]$$\xymatrix{1\circ  & 2\circ \ar@/^/[rrrrr]^{\mathrm{1}} & 3\circ & 4\circ & 5\circ & 6\circ & 7\circ\ar@/_/[lllll] &
8\circ}$$\caption{$G_2$}\label{Fig:GHZ3q2}\end{figure}

\begin{figure}[h!]$$\xymatrix{1\circ  & 2\circ & 3\circ\ar@/^/[rrr]^{\mathrm{1}} & 4\circ & 5\circ & 6\circ\ar@/_/[lll] & 7\circ &
8\circ}$$\caption{$G_3$}\label{Fig:GHZ3q3}\end{figure}

\begin{figure}[h!]$$\xymatrix{1\circ  & 2\circ & 3\circ & 4\circ\ar@/^/[r]^{\mathrm{1}} & 5\circ\ar@/_/[l] & 6\circ & 7\circ &
8\circ}$$\caption{$G_4$}\label{Fig:GHZ3q4}\end{figure} and the density matrices associated with $G_1, G_2, G_3, G_4$ are given
$\sigma(G_i)=\frac{K(G_i)}{\tr(K(G_i))}$ where $K(G_i)\in\{L(G_i), Q(G_i)\}$, and $\sigma(G_1)=\left| \psi  \right\rangle _{123}^{(1),(2)}  \left\langle \psi
\right|_{123}^{(1),(2)}$, $\sigma(G_2)=\left| \psi  \right\rangle _{123}^{(3),(4)}  \left\langle \psi  \right|_{123}^{(3),(4)}$, $\sigma(G_3)=\left| \psi  \right\rangle
_{123}^{(5),(6)}  \left\langle \psi  \right|_{123}^{(5),(6)}$, $\sigma(G_4)=\left| \psi  \right\rangle _{123}^{(7),(8)}  \left\langle \psi  \right|_{123}^{(7),(8)}$.  \par
The general three-qubit $W$ state is given as $\left|\psi\right\rangle_{123}^{W}=a|001\rangle +b|010\rangle +c|100\rangle$
where $|a|^2+|b|^2+|c|^2=1$. The graph representation of $\left|\psi\right\rangle_{123}^{W}$ is
given in Figure \ref{Fig:wstate2}.

\begin{figure}[h!]$$\xymatrix{\ar@(ul,dl)_{|a|(|a|-|b|-|c|)}2\ar[r]^{a\overline{b}} \ar[d]_{a\overline{c}} & 4\ar@(ur,dr)^{|b|(|b|-|a|-|c|)} \\
\ar@(dl,ul)^{|c|(|c|-|a|-|b|)}5\ar[ru]_{c\overline{b}} & 1\circ & 3\circ & 6\circ & 7\circ & 8\circ}$$\caption{G}\label{Fig:wstate2}
\end{figure}
The density matrix for the $W$ class of states can be expressed as $\sigma(G)=\frac{L_-(G)}{\tr(L_-(G))}$. For a specific case
where $a=b=c=\frac{1}{\sqrt{3}}$, Figure 17 represents the graphical representation for a standard $W$ state.
\begin{figure}[h!]
$$\xymatrix{\ar@(ul,dl)_{-1}2\ar@{-}[r]^{\mathrm{1}} \ar@{-}[d]_{\mathrm{1}} & 4\ar@(ur,dr)^{-} & 1\circ & 3\circ & 6\circ & 7\circ & 8\circ \\ \ar@(dl,ul)^{-1}5\ar@{-}[ru]_{\mathrm{1}}}$$\caption{G}\label{Fig:wstate}
\end{figure}
It is evident that the graphs for GHZ and W classes are completely distinct from each other. Therefore, using our approach one can easily identify whether a given three-qubit
state belongs to a GHZ class or W class. \par
Similar to $G_i,i =1,\hdots, 4$ graphs with $8$ vertices, graphs with $16$ vertices can be produced which will provide the GHZ states with $4$-qubits. By similar graphs, we mean
graphs with $16$ nodes, one edge which connects the $i$th and $(17-i)$th vertices, $i=1, \hdots, 8$ having edge weight $1$. Similarly one can also obtain the graph for a four
qubit $W$ state.
\item \textit {Cluster and Chi states}: The four qubit cluster \cite{Briegel} and Chi \cite{Chua} states are given by
$\left|\psi\right\rangle_{1234}=\frac{1}{2}(|0000\rangle) + |0101\rangle + |1010\rangle - |1111\rangle)$ and
$\left|\phi\right\rangle_{1234}=\frac{1}{2}(|0000\rangle) + |0101\rangle + |1011\rangle - |1110\rangle)$, respectively. The
corresponding graphs for these two states are given in Figures \ref{Fig:cluster} and  \ref{Fig:chi}, respectively.
\begin{figure}[h!]
$$\xymatrix{\ar@(ul,dl)_{-2}1\circ \ar@{-}[r]^{\mathrm{1}} \ar@{-}[d]_{\mathrm{-1}} \ar@{-}[rd]^{\mathrm{1}} & 6\circ \ar@(ur,dr)^{-2}\ar@{-}[d]^{\mathrm{-1}} & 2\circ & 3\circ & 4\circ & 5\circ & 7\circ & 8\circ \\ \ar@(dl,ul)^{-2}11\circ \ar@{-}[r]_{\mathrm{-1}}\ar@{-}[ru]_{\mathrm{-1}} & \ar@(dr,ur)_{-2}16\circ & 9\circ & 10\circ & 12\circ & 13\circ &14\circ &15\circ} $$\caption{G}\label{Fig:cluster}
\end{figure}
\begin{figure}[h!]
$$\xymatrix{\ar@(ul,dl)_{-2}1\circ \ar@{-}[r]^{\mathrm{1}} \ar@{-}[d]_{\mathrm{1}} \ar@{-}[rd]^{\mathrm{1}} & 6\circ \ar@(ur,dr)^{-2}\ar@{-}[d]^{\mathrm{-1}} & 2\circ & 3\circ & 4\circ & 5\circ & 7\circ & 8\circ \\ \ar@(dl,ul)^{-2}12\circ \ar@{-}[r]_{\mathrm{-1}}\ar@{-}[ru]_{\mathrm{-1}} & \ar@(dr,ur)_{-2}15\circ & 9\circ & 10\circ & 11\circ & 13\circ &14\circ &16\circ} $$\caption{G}\label{Fig:chi}
\end{figure}
The two states are different as evident from the edge weights. Similarly, the density matrices corresponding to Cluster and Chi
states are represented by
$\sigma(G_{i})=\frac{L_-(G_{i})}{\tr(L_-(G_{i}))}$ where $\sigma(G_{1})=\left| \psi  \right\rangle _{1234} \left\langle \psi  \right|_{1234}$, and $\sigma(G_{2})=\left| \phi
\right\rangle _{1234} \left\langle \phi  \right|_{1234}$.

\item \textit {Brown State}: Consider the graph $G$ with $32$ vertices given in Figure (20). The isolated vertices are not shown in the
graph and the weights of the edges and loops are as given below.
\begin{eqnarray}
w_{ij}=\left\{
           \begin{array}{ll}
             -1, & \hbox{$i=1, j=4, 15 $} \\
             1, & \hbox{$i=1, j=14, 21, 24, 26, 27$} \\
             -1, & \hbox{$i=4, j=14, 21, 24, 26, 27$} \\
             1, & \hbox{$i=4, j=15 $} \\
 -1, & \hbox{$i=14, j=15 $} \\
 1, & \hbox{$i=14, j=21, 24, 26, 27$} \\
- 1, & \hbox{$i=15, j=21, 24, 26, 27 $} \\
 1, & \hbox{$i=21, j=24, 26, 27 $}\\
 1, & \hbox{$i=24, j=26, 27$}\\
 1, & \hbox{$i=26, j=27 $}\\
 -6, & \hbox{$i=j, i=1,4, 14,15,21,24,26,27. $}
           \end{array}
         \right.
\end{eqnarray}
The graph in Figure (20) represents a five qubit Brown state \cite{Brown}, namely
\begin{eqnarray}
\label{Bstate}
\left| \psi\right\rangle^{12345}  &= &\frac{1}{2\sqrt{2}} \left[\left| 00000 \right\rangle - \left| 00011 \right\rangle +\left| 01101 \right\rangle -\left| 01110 \right\rangle
+\left| 10100 \right\rangle +\left| 10111 \right\rangle +\left| 11001 \right\rangle\right. \nonumber \\ &+& \left.\left| 11010\right\rangle\right]
\end{eqnarray}
In comparison to other non-equivalent classes of five-qubit entangled states, Brown states are said to be more entangled. The reason is evident from the property that all the
bipartitions of Brown states are maximally mixed which is not the case with GHZ, Cluster or Chi type of states. \par
The density matrix associated with the graph for Brown state is \begin{small}\beano \sigma(G) &=& \frac{1}{\tr(L_-(G))}L_-(G) = \left| \psi  \right\rangle _{12345} \left\langle \psi  \right|_{12345}. \eeano\end{small}
\begin{figure}[h!]
$$\xymatrix{ & \ar@(u,l)1\circ\ar@{-}[rr] \ar@{-}[dl] \ar@{-}[drrr] \ar@{-}[ddl] \ar@{-}[ddd]  \ar@{-}[dddrr] \ar@{-}[ddrrr]& & \ar@(u,r)4\circ\ar@{-}[dr]\ar@{-}[dlll]
\ar@{-}[ddr] \ar@{-}[ddd] \ar@{-}[dddll] \\ \ar@(ul,dl)14\circ \ar@{-}[rrrr] \ar@{-}[d] \ar@{-}[drrrr] \ar@{-}[ddrrr] \ar@{-}[ddr]& & & & \ar@(ur,dr)15\circ\ar@{-}[d] \ar@{-}[ddl]
\ar@{-}[ddlll] \ar@{-}[dllll]\\ \ar@(ul,dl)21 \circ \ar@{-}[rrrr] \ar@{-}[dr] \ar@{-}[drrr] \ar@{-}[uurrr]& & & & \ar@(ur,dr)24 \circ\ar@{-}[dl] \ar@{-}[dlll]\\ & \ar@(d,l)26\circ
\ar@{-}[rr] & & \ar@(d,r)27\circ}$$\caption{G}\label{Fig:Bstate}
\end{figure}
\item {\it 5-qubit Chi state}: The five-qubit Chi state \cite{Man} can be represented as
\begin{eqnarray}
\label{chi5}
\left| \phi\right\rangle_{12345}  &= &\frac{1}{2}\left[|00000\rangle +|00111\rangle + |01010\rangle - |01101\rangle - |10011\rangle + |10100\rangle + |11001\rangle \right.
\nonumber \\ &+& \left.|11110\rangle\right].
\end{eqnarray}
The graph and weights of edges for the five-qubit Chi state are given by Figure (21) and Eq. (14), respectively.
\begin{figure}[h!]$$\xymatrix{ & \ar@(u,l)1\circ\ar@{-}[rr] \ar@{-}[dl] \ar@{-}[drrr] \ar@{-}[ddl] \ar@{-}[ddd]  \ar@{-}[dddrr] \ar@{-}[ddrrr]& & \ar@(u,r)8\circ\ar@{-}[dr]\ar@{-}[dlll] \ar@{-}[ddr] \ar@{-}[ddd] \ar@{-}[dddll] \\ \ar@(ul,dl)11\circ \ar@{-}[rrrr] \ar@{-}[d] \ar@{-}[drrrr] \ar@{-}[ddrrr] \ar@{-}[ddr]& & & & \ar@(ur,dr)14\circ\ar@{-}[d] \ar@{-}[ddl] \ar@{-}[ddlll] \ar@{-}[dllll]\\ \ar@(ul,dl)20 \circ \ar@{-}[rrrr] \ar@{-}[dr] \ar@{-}[drrr] \ar@{-}[uurrr]& & & & \ar@(ur,dr)21 \circ\ar@{-}[dl] \ar@{-}[dlll]\\ & \ar@(d,l)26\circ \ar@{-}[rr] & & \ar@(d,r)31\circ}$$\caption{G}\label{Fig:Bstate}\end{figure}
\begin{eqnarray}
w_{ij}=\left\{
           \begin{array}{ll}
             -1, & \hbox{$i=1, j=14, 20 $} \\
             1, & \hbox{$i=1, j=8, 11, 21, 26, 31$} \\
             -1, & \hbox{$i=8, j=14, 20$} \\
             1, & \hbox{$i=8, j= 11, 21 , 26, 31 $} \\
 -1, & \hbox{$i=11, j=14,20 $} \\
 1, & \hbox{$i=11, j=21, 26, 31$} \\
- 1, & \hbox{$i=14, j= 21,26,31 $} \\
 1, & \hbox{$i=14, j=20 $}\\
 -1, & \hbox{$i=20, j=21,26, 31$}\\
 1, & \hbox{$i=26, j=31 $}\\
 -6, & \hbox{$i=j, i=1,4, 14,15,21,24,26,27. $}
           \end{array}
         \right.
\end{eqnarray}
The density matrix, therefore, can be given as $\sigma(G) = \frac{1}{\tr(L_-(G))}L_-(G) = \left| \phi  \right\rangle _{12345} \left\langle \phi  \right|_{12345}$. \par
Although the difference between Brown and Chi states can be characterized from the edge weights, for a meaningful
classification of such states using a graph theoretical approach,
one needs to quantify a graph theoretic measure for entanglement. Such a measure will classify quantum states in different
classes and provide deeper physical insight into the complex
nature of multiqubit entanglement.




\end{enumerate}


\begin{remark}
Observe that in the graph representation of entangled pure states mentioned above, all the existing weighted edges are clustered in a completely connected subgraph of the original graph. Further, the weight of the loops attached at each of the vertices of the subgraph is $-(m-2)$ where $m$ is the number of vertices involved in the complete subgraph.
\end{remark}


\section{Conclusion}

 We define combinatorial, signless and signed Laplacian matrices associated with a weighted digraph having complex edge weights with or without loops. We determine
the connection between the existence of zero Laplacian eigenvalues of a weighted digraph and the topological structure of the graph. Using these Laplacian matrices, we define density
matrices corresponding to a weighted digraph. We have classified graphs which represent pure and mixed density matrices of
quantum states by using the topological structure of the graphs. This work initiates a number of directions to the combinatoric visualization of quantum mechanical phenomena. Some
 of them are listed below.

\begin{enumerate}
\item
A state is called separable if the density matrix, $$\rho = \sum_i p_i \rho_i^{(A)} \otimes \rho_i^{(B)}.$$
Here, $\rho_i^{(A)}$ and $\rho_i^{(B)}$ denotes density matrix of subsystems $A$ and $B$. We denote tensor product of matrices by $\otimes$.
The state is entangled otherwise.  We have deduced the graphs for several well-known entangled pure states. We have demonstrated that the three-qubit
entangled systems can be classified into GHZ and W class using a graph theoretic approach. A criteria of separability states represented by the Laplacian of simple graphs has been developed in
\cite{WChiWa}. A combinatorial operation has also been introduced for density matrices defined by Laplacian matrices associated with simple graphs in \cite{DuAdBaSr}
that act as an entanglement generator for mixed states arising from partially symmetric graphs. These works introduce new results for the
separability of density matrices corresponding to weighted digraphs.

\item
In order to develop further insight into the entanglement properties of multiqubit systems, it would be interesting to define a graph theoretic measure for
quantification and classification of entanglement in such systems.

\item
Recently, local unitary transformations on a density matrix obtained by signless Laplacian matrix associated with a simple graph has been established as a combinatorial
operation which is known as switching of a graph in \cite{DuAdBa}. This work sheds further light to the problem of unitary equivalence and state classification for the
states related to weighted digraphs.
\end{enumerate}

This work is, we hope, a contribution towards a new direction in the field of quantum information.

\end{document}